\documentclass[runningheads]{llncs}
\usepackage{footmisc}
\usepackage{graphicx}
\usepackage{booktabs}
\usepackage{cite}

\usepackage{hyperref}

\makeatletter
\newcommand{\printfnsymbol}[1]{%
  \textsuperscript{\@fnsymbol{#1}}%
}
\makeatother

\begin{document}
%
\title{Enhanced generative adversarial network for 3D brain MRI super-resolution}
\titlerunning{EGN for brain MRI SR}

\author{Jiancong Wang\inst{1} \thanks{equal contribution}
\and
Yuhua Chen\inst{2} \printfnsymbol{1}   
\and
Yifan Wu\inst{1} 
\and
Jianbo Shi\inst{3} 
\and
James Gee\inst{1,3} 
}


%
%
\institute{Penn Image Computing and Science Laboratory, University of Pennsylvania, Philadelphia, PA 19104, USA
\and
Department of Bioengineering, University of California, Los Angeles, CA 90095, USA
\and
Department of Computer and Information Science, University of Pennsylvania, Philadelphia, PA 19104, USA
}


\maketitle              

\begin{abstract}
Single image super-resolution (SISR) reconstruction for magnetic resonance imaging (MRI) has generated significant interest because of its potential to not only speed up imaging but to improve quantitative processing and analysis of available image data. 
Generative Adversarial Networks (GAN) have proven to perform well in recovering image texture detail, and many variants have therefore been proposed for SISR.
In this work, we develop an enhancement to tackle GAN-based 3D SISR by introducing a new residual-in-residual dense block (RRDG) generator that is both memory efficient and achieves state-of-the-art performance in terms of PSNR (Peak Signal to Noise Ratio), SSIM (Structural Similarity) and NRMSE (Normalized Root Mean Squared Error) metrics. We also introduce a patch GAN discriminator with improved convergence behavior to better model brain image texture. We proposed a novel the anatomical fidelity evaluation of the results using a pre-trained brain parcellation network. Finally, these developments are combined through a simple and efficient method to balance between image and texture quality in the final output.





\end{abstract}

\section{Introduction}
High spatial resolution (HR) structural MRI provides fine-grain anatomical information and make accurate quantitative image analysis feasible, benefiting clinical diagnosis and precise decision making. However, it not only requires expensive equipment but also longer scan time, which introduces both acquisition challenges~\cite{plenge2012super} and potentially limits clinical accessibility in situations where only short scans are feasible. Single image super-resolution (SISR) reconstruction for MRI has generated interest because HR images may potentially be derived from rapidly generated low resolution images obtained by subsampling k-space in MRI.


Many solutions using 3D CNNs (convolutional neural networks) have been proposed for the medical imaging SISR problem~\cite{chen2018efficient, pham2017brain, zhao2018self}. Sanchez et al.~\cite{sanchez2018brain} adapted the standard super-resolution GAN (SRGAN) \cite{ledig2017photo} framework for brain image super-resolution. Zhao et al.~\cite{zhao2018self} developed an improved super-resolution residial network (SRResNet) \cite{ledig2017photo} for axial slice super-resolution. Chen et al.~\cite{chen2018efficient} proposed a multi-layer DenseNet~\cite{huang2017densely} based network for fast and efficient inference and WGAN training~\cite{gulrajani2017improved} for realistic texture recovery. 
A challenge with these efforts~\cite{chen2018efficient, sanchez2018brain} is that the applied discriminator requires long convergence times because of numerical instability.


The developments here applies generally to medical imagery but, for concreteness, are instantiated for SISR in T1 MR images of the brain, as in \cite{chen2018efficient}. Assuming a k-space subsampled low resolution image as input, we propose to extend the GAN model for SISR as follows: 1) a new generator backbone is developed based on 3D residual-in-residual dense block~\cite{wang2018esrgan}, that is both high performance and memory efficient; 2) a patch GAN~\cite{isola2017image} discriminator that is more numerically stable; 3) an efficient blending approach is implemented to trade off between PSNR and GAN oriented models to generate the final output; and 4) a novel metric quantifying anatomical fidelity is introduced using a pre-trained brain parcellation network. 






\section{Methods}



\begin{figure}
\begin{center}
\includegraphics[width=0.8\textwidth]{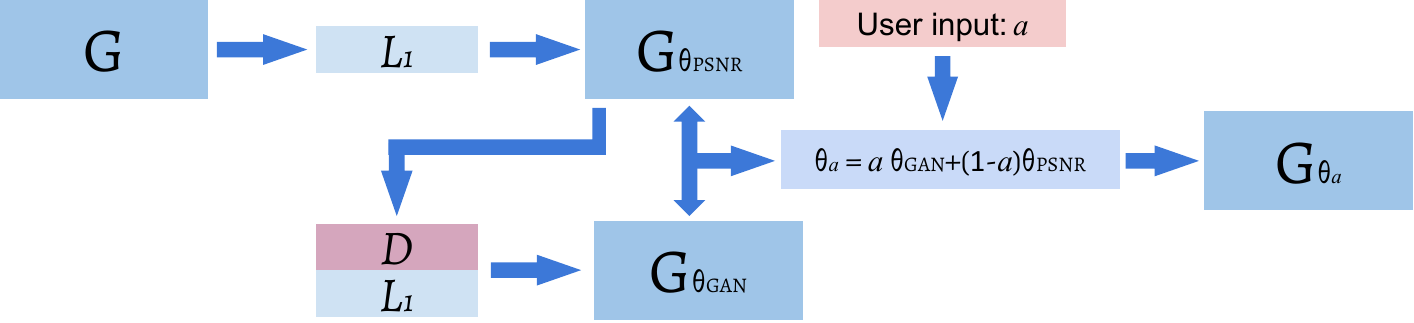}
\caption{Model training and blending pipeline. }
\label{train_and_blend}
\end{center}
\end{figure}

The overall pipeline is illustrated in~Fig.\ref{train_and_blend}.
The generator is trained with $L_1$ loss to obtain our PSNR oriented model, and is fine-tuned using a patch GAN discriminator to be our GAN oriented model. The former model is optimized with respect to conventional similarity metric but ignores textural fidelity, while the latter GAN model incorporates but at the expense of potentially introducing artifact into the output. A blending parameter $\alpha$ permits trade off between the two models in the final generated image. 

\subsection{Memory efficient residual-in-residual dense block generator (RRDG)}
In SISR task with GAN framework, the network architecture of the generator is of paramount importance of generated image quality. Ledig et al.~\cite{ledig2017photo} introduced a residual network~\cite{he2016deep} for SISR and Zhang et al. ~\cite{zhang2018residual} extends the idea with the residual in residual connection. Chen et al.~\cite{chen2018efficient}, on the other hand, adapted the dense net to for the SISR task. 
Combining the residual connection and dense connection, Wang et al.~ \cite{wang2018esrgan} proposed a hybrid of residual and dense connections, termed residual-in-residual dense block (RRDB), to replace the basic residual block in SRResNet. It achieved the state-of-the-art performance in the PIRM2018-SR competition \cite{ignatov2018pirm}. In this work, we extended the RRDB network to 3D for brain MRI SISR. We called it residual-in-residual-dense-block generator(RRDG). 

\begin{figure}
\includegraphics[width=\textwidth]{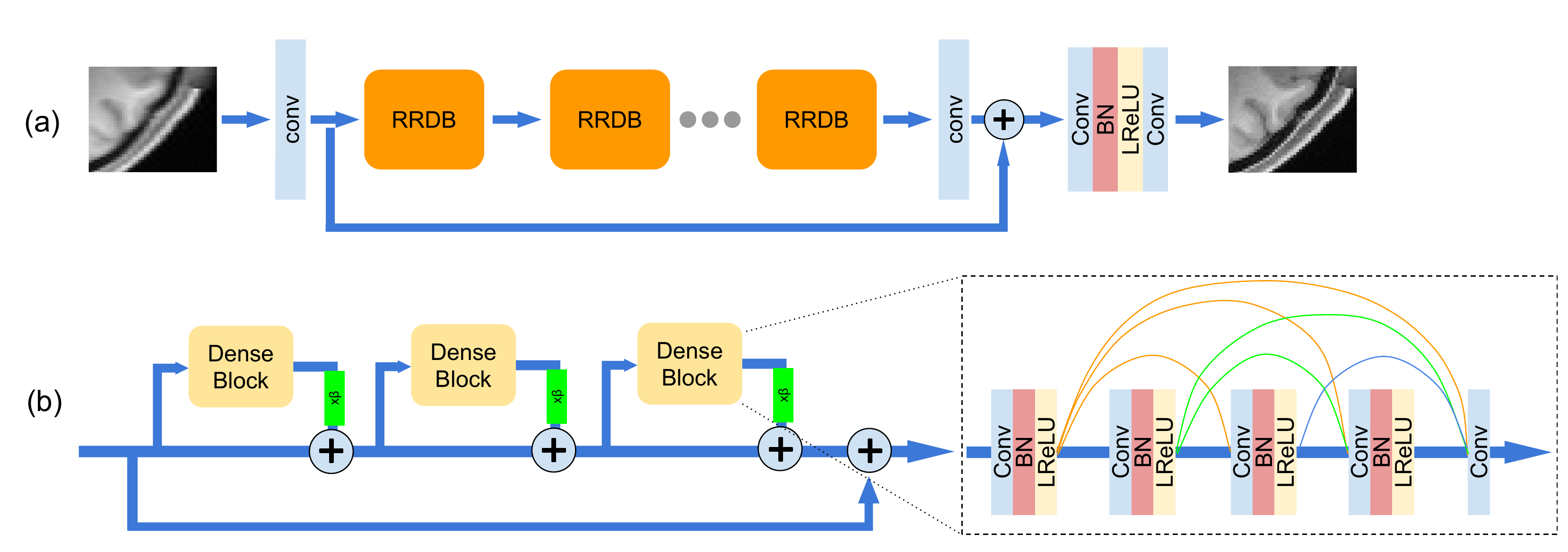}
\caption{Architecture of the proposed RRDG network and RRDB. Like SRResNet~\cite{isola2017image}, RRDG consists of a global residual connection and consecutive basic blocks, except basic resblocks are replaced by RRDB. Within each RRDB, three consecutive dense blocks are chained by a weighted residual connection and a block level residual connection.}
\label{rrdb}
\end{figure}

The proposed RRDG network is shown in Figs.~\ref{rrdb}. 
Note that in each RRDB the feature is residual summed; therefore, the overall number of features $n_f$ is unchanged throughout all blocks. Compared to mDCSRN proposed in~\cite{chen2018efficient}, which densely accumulates features globally and grows wider as the network gets deeper, the RRDG remains narrow. This memory efficient aspect of RRDB is an important advantage for 3D applications as in the current brain MRI SISR task. 


\subsection{Patch GAN discriminator}

\begin{figure}
\includegraphics[width=\textwidth]{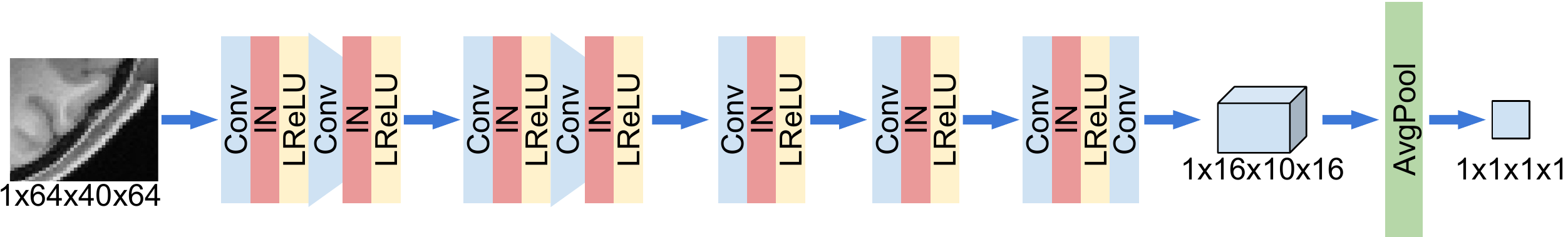}
\caption{Architecture of the discriminator, which is a VGG-style feed forward network~\cite{simonyan2014very}, consisting of 2 strided convolution blocks for down-sampling followed by plain convolution-layer norm-leakyReLU layers. The outputs are globally spatial pooled to produce a single number.} 
\label{patch_gan}
\end{figure}

Our discriminator follows the work in~\cite{chen2018efficient, ledig2017photo}, as shown in Fig.~\ref{patch_gan}. However, the latter discriminators end with two dense layers, and during training, we observed unstable discriminator output due to large $|w|$ and $||w||$, where $w$ is the weight of dense layer. This instability slows training convergence. We also observed that the rows of $w$ of these layers are highly redundant. We chose therefore to replace the final dense layers with a single $1\times1\times1$ convolution and a global mean pooling layer, as in patch GAN~\cite{isola2017image}. 
Furthermore, by removing the final two strided convolutions, we reduced the receptive field of our discriminator and facilitate discrimination of local texture that in turn produces more accurate texture synthesis. 

WGAN-GP training~\cite{gulrajani2017improved} was used in our implementation for its stability; specifically, replacing batch norm with instance norm, adding a gradient penalty to the discriminator for randomly interpolated $\{I_{hr},I_{sr}\}$ (high resolution ground truth, super-resolution) inputs, and scheduling the discriminator to always run ahead of the generator. 


The loss functions $L_{G}$ and $L_{D}$ for the generator and discriminator, respectively, are defined as follows:

\begin{equation}
L_{G} = L_1(I_{sr}, I_{hr}) + \lambda_D D(I_{sr}),
\label{gloss}
\end{equation}
\noindent
where $L_1$ is the element-wise $L_1$ loss, $D$ is the discriminator and $\lambda_D$ being the weighting factor between the two terms; 
\begin{equation}
L_{D} = D(I_{hr}) - D(I_{sr}) + \lambda_{g}  || \dot{D}( \gamma I_{sr} + (1-\gamma)  I_{hr}) ||_2, \gamma \sim (0, 1),
\label{dloss}
\end{equation}
\noindent
where $\dot{D}$ is the derivative of the discriminator, $\lambda_{g}$ is the weighting factor added on the gradient penalty, and $\gamma$ is a random number drawn from a uniform distribution $(0, 1)$ for each batch.
With our modifications to the discriminator and use of the WGAN training procedure, model training consistently converges within 10 epochs. 



\subsection{Enhancing texture realism through blending of PSNR oriented model and GAN model}
GAN model qualitatively improves texture fidelity, at the cost, however, of reduced image quality of the generated images as assessed by conventional metrics~\cite{isola2017image}. For human perceptual tasks, this image `degradation' may be acceptable (even desirable), whereas for machine processing, the lower image quality may negatively impact algorithm performance. One solution that we develop here, following the work in~\cite{wang2018esrgan}, is to permit model blending through linear combination of model parameters for our PSNR and GAN models. Let $\theta_G$ represent parameters of the generator incorporating both PSNR and GAN models, and $\alpha$, a user defined linear blending weight. We define the blended model as:  

\begin{equation}
\theta_{G}^{\alpha} = \alpha \theta_{G}^{PSNR} + (1-\alpha) \theta_{G}^{GAN}.
\label{model_blending}
\end{equation}


\subsection{HighRes3D net for evaluating anatomical fidelity}

As mentioned in \cite{borji2019pros}, although structure similarity metrics such as SSIM are broadly used, image generation lacks of perceptually meaningful measures. Specially in this MRI super-resolution task, it is critical to measure the quality of synthesized images from a clinical usage perspective. Here, we propose a novel metric to assess anatomical accuracy. More specifically, we use HighRes3D \cite{li2017compactness}, a pre-trained brain parcellation network, to evaluate the distance between HR images and synthesized images in the anatomical distinctive space and calculate the dice scores. 

HighRes3D consists of a primary convolution, 3 residual blocks, and 2 final convolutions. We define our metric $M_h$ as a weighted sum of $L_1$ distance of features from these 6 layers between SR and HR images:
\begin{equation}
M_{h} = \sum_{f \in features} w_f L_1 (f (I_{sr}), f (I_{hr}) ).
\label{hres}
\end{equation}
where $f$ are features output from the 6 layers. $w_f$ are weights of each of the feature layers, chosen to normalized scale of each feature layers so that each feature layer contribute equally. 



\section{Experimental Results}
\subsection{Methodology}
Ground truth images were obtained from the Human Connectome Project (HCP). Specifically, it includes 1,113 3D T1 MR images from 1,200 healthy young subjects on Siemens 3T platform and 32-channel head coil. Imaging parameters were: 3D MPRAGE sequence, 0.7 mm$^3$ isotropic voxels, $320\times320\times256$ matrix size. HCP images were downsampled to 1 mm$^3$ resolution using spline interpolation for our SISR experiments. Low resolution versions of these images were created by further downsampling the resolution in coronal and sagittal planes by one half in k-space following the procedure in~\cite{chen2018efficient}. 


The same number of splits was used as in~\cite{chen2018efficient}; specifically, 780 for training, 111 for validation, 111 for evaluation, and 111 for testing. Results are reported on the test set, which was not used in model training or parameter optimization. 
We used a patch size of 64$\times$40$\times$64 as input due to GPU memory constraints, and cropped 3 voxels around the boundary of the output to avoid discontinuity around edges, resulting in 58$\times$32$\times$58 output patches. The complete output image is assembled by stitching together non-overlapping output patches.
We implemented our model in PyTorch 1.0 and trained the model on a workstation with 4 GTX 1080 Ti GPUs.

For the generator, we implemented the dense accumulation within each RRDB with gradient check-pointing~\cite{pleiss2017memory} for memory efficiency. Our adapted RRDG network contains $n_c = 4$ RRDB and has residual feature width of $n_f = 48$. Within each dense block we set the dense growth rate $k = 12$. These values were established based on hyper-parameter search. We used batch normalization in the RRDB, and kept the scaling factor $\beta = 0.2$ and activation as leaky rectified linear unit as in the original implementation. 





\subsection{Results}
\begin{figure}
\includegraphics[width=\textwidth]{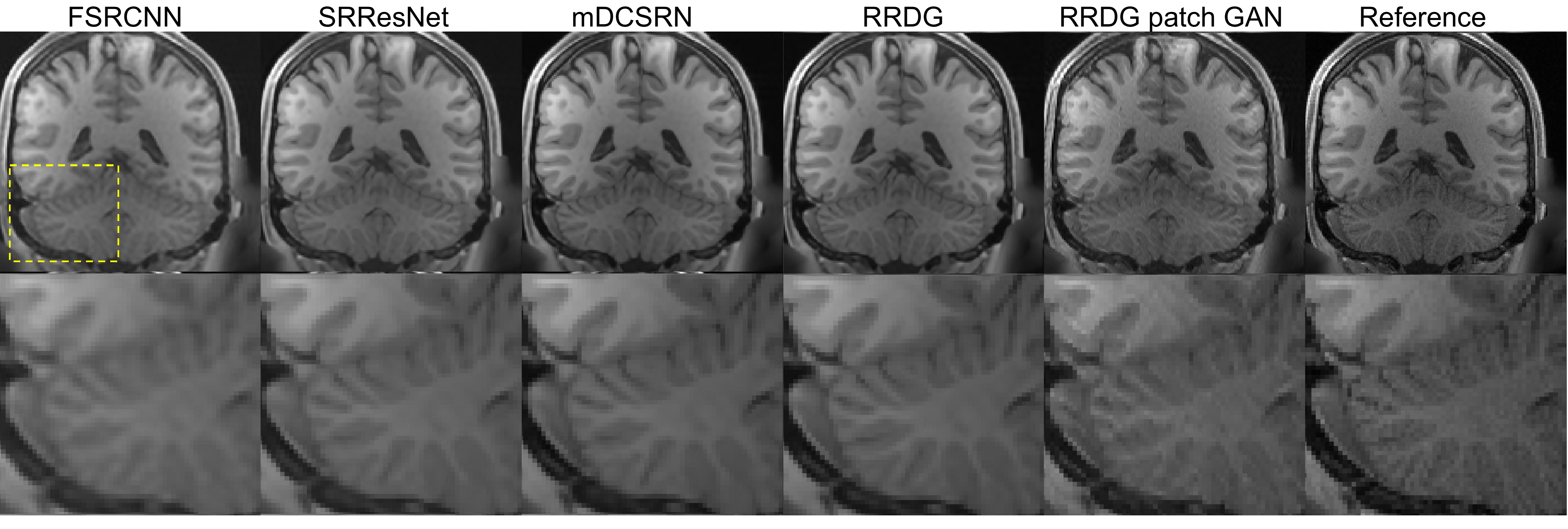}
\caption{Left to right: Super-resolution output from FSRCNN, SRResNet, mDCSRN, RRDB, RRDB with patch GAN training, and ground truth. Bottom row is magnified portion of the same image region across the different SISR outputs.}
\label{comparison}
\end{figure}

\noindent
Our proposed RRDG and its patch GAN augmented variant were evaluated against state-of-the-art FSRCNN, SRResNet, and mDCSRN models for SISR reconstruction. The FSRCNN and SRResNet are adapted to 3D directly. For the mDCSRN, we used the b8u4 configuration, 8 dense blocks with 4 dense layers within each block. Example results are illustrated in Fig.~\ref{comparison}. Visually the output from RRDB is most sharp and the GAN oriented model successfully recover qualitatively more realistic brain image texture.

For quantitative comparison, we used SSIM, PSNR, NRMSE metrics, the anatomical fidelity metric $M_h$, and average dice score on 160 tissue types between HighRes3DNet segmentation generated from SR images and HR images. We normalized the image using N4 bias correction~\cite{tustison2010n4itk} before applying the proposed anatomical fidelity metric $M_h$ as HighRes3DNet~\cite{li2017compactness} expects normalized input.

As can be seen in Table~\ref{benchmark}, RRDG achieves the best performance with statistically significant improvement across all evaluation metrics. We performed two-tailed pair-wise t-test and for all metrics, $p < 1.1e-53$. For fair comparison, the total number of parameters and run time on single image are shown.


\begin{table}\centering
\caption{SISR image quality and anatomical fidelity assessment}
\resizebox{\textwidth}{!}{
\begin{tabular}{@{}r cccr cccr cccr ccc@{}}\toprule
& \multicolumn{3}{c}{3D FSRCNN \cite{dong2016accelerating}} && \multicolumn{3}{c}{3D SRResNet \cite{ledig2017photo}} && \multicolumn{3}{c}{mDCSRN(b8u4)\cite{chen2018efficient}} && \multicolumn{3}{c}{RRDG (ours)}\\
\cmidrule{2-4} \cmidrule{6-8} \cmidrule{10-12} \cmidrule{14-16}
& SSIM & PSNR & NRMSE && SSIM & PSNR & NRMSE && SSIM & PSNR & NRMSE && SSIM & PSNR & NRMSE\\ 
\midrule
mean & 0.9282 & 33.83 & 0.1138 && 0.9399 & 34.06 & 0.1104 && 0.9485 & 35.38 & 0.0954 && \textbf{0.9558} & \textbf{36.39} & \textbf{0.0846}\\
std & 0.0068 & 1.0376 & 0.0046 && 0.0068 & 0.9775 & 0.0055 && 0.0059 & 1.0624 & 0.0042 && 0.0066 & 1.0370 & 0.0045\\
\toprule
$M_h$ & \multicolumn{3}{c}{7.3847} && \multicolumn{3}{c}{6.3489} && \multicolumn{3}{c}{5.4446} && \multicolumn{3}{c}{\textbf{4.8217}}\\

Dice & \multicolumn{3}{c}{0.8677} && \multicolumn{3}{c}{0.9048} && \multicolumn{3}{c}{0.9153} && \multicolumn{3}{c}{\textbf{0.9249}}\\

\#param & \multicolumn{3}{c}{64893} && \multicolumn{3}{c}{2004620} && \multicolumn{3}{c}{625969} && \multicolumn{3}{c}{2650129}\\
Time(s) & \multicolumn{3}{c}{7.4} && \multicolumn{3}{c}{80.3} && \multicolumn{3}{c}{23.0} && \multicolumn{3}{c}{33.9}\\

\bottomrule
\end{tabular}
}
\label{benchmark}
\end{table}

\begin{figure}
\includegraphics[width=\textwidth]{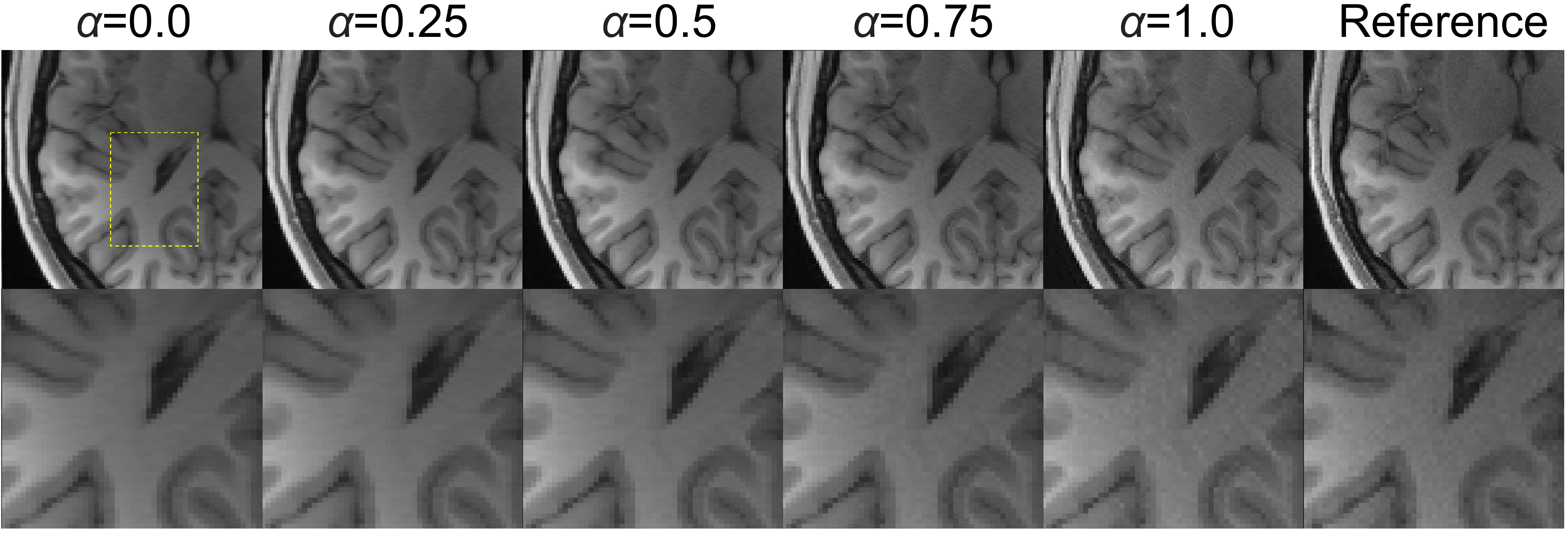}
\caption{Sample image appearance as a function of blending between GAN oriented model ($\alpha = 1$) and PSNR oriented model ($\alpha = 0$), compared with ground truth.} 
\label{blending_result}
\end{figure}

Fig.~\ref{blending_result} illustrates the effect of model blending on the generated output image. The output varies smoothly with the interpolation factor $\alpha$, allowing controllable trade off between reconstruction image and texture fidelity. 
Compared to blending model output images, mixing model parameters yields smoother results~\cite{wang2018esrgan}, and, equally important, is also more efficient since blending parameters requires minor computations and the generator only run once, while model output blending requires the generator to be run twice on two set of parameters. 

\section{Discussion}



In this work, we investigated enhancements to CNN-based solutions to 3D brain MRI super-resolution. The RRDG was shown to exhibit superior performance against the state-of-the-art, and amenable to memory optimization to make possible efficient training. We also introduced a patch GAN discriminator to better model brain image texture and optimized the network for more stable convergence behavior.
A novel metric is developed to assess anatomical accuracy of the reconstructions. These developments can be combined to balance between image and texture fidelity in the final output.


There are many directions for ongoing and future work. Hardware limitations precluded a more thorough ablation study to interrogate model architectures, which remains a major focus of our current work. Additional comparisons to existing and new work in the field, including, for example, \cite{simonyan2014very}, is another priority. An open challenge is artifacts in patch GAN reconstructions, and new solution possibilities include improved generator architectures and the incorporation of domain information such as brain image segmentations~\cite{wang2018recovering}. 
Similarly, the proposed mixture model for balancing super-resolution image and texture quality is not fully developed. We aim to conduct an expert observer study to determine radiologist-optimized blended image output, and compare these against the default networks in a task-based evaluation of brain segmentation accuracy. 




\bibliographystyle{splncs04}
\bibliography{mybib.bib}

\begin{thebibliography}{10}
\providecommand{\url}[1]{\texttt{#1}}
\providecommand{\urlprefix}{URL }
\providecommand{\doi}[1]{https://doi.org/#1}

\bibitem{borji2019pros}
Borji, A.: Pros and cons of gan evaluation measures. Computer Vision and Image
  Understanding  \textbf{179},  41--65 (2019)

\bibitem{chen2018efficient}
Chen, Y., Shi, F., Christodoulou, A.G., Xie, Y., Zhou, Z., Li, D.: Efficient
  and accurate {MRI} super-resolution using a generative adversarial network
  and 3{D} multi-{L}evel densely connected network. In: MICCAI (2018)

\bibitem{dong2016accelerating}
Dong, C., Loy, C.C., Tang, X.: Accelerating the super-resolution convolutional
  neural network. In: ECCV (2016)

\bibitem{gulrajani2017improved}
Gulrajani, I., Ahmed, F., Arjovsky, M., Dumoulin, V., Courville, A.C.: Improved
  training of wasserstein gans. In: NeurIPS (2017)

\bibitem{he2016deep}
He, K., Zhang, X., Ren, S., Sun, J.: Deep residual learning for image
  recognition. In: CVPR (2016)

\bibitem{huang2017densely}
Huang, G., Liu, Z., Van Der~Maaten, L., Weinberger, K.Q.: Densely connected
  convolutional networks. In: CVPR (2017)

\bibitem{ignatov2018pirm}
Ignatov, A., Timofte, R., Van~Vu, T., Luu, T.M., Pham, T.X., Van~Nguyen, C.,
  Kim, Y., Choi, J.S., Kim, M., Huang, J., et~al.: Pirm challenge on perceptual
  image enhancement on smartphones: report. In: European Conference on Computer
  Vision. pp. 315--333. Springer (2018)

\bibitem{isola2017image}
Isola, P., Zhu, J.Y., Zhou, T., Efros, A.A.: Image-to-image translation with
  conditional adversarial networks. In: CVPR (2017)

\bibitem{ledig2017photo}
Ledig, C., Theis, L., Husz{\'a}r, F., Caballero, J., Cunningham, A., Acosta,
  A., Aitken, A., Tejani, A., Totz, J., Wang, Z., et~al.: Photo-realistic
  single image super-resolution using a generative adversarial network. In:
  CVPR (2017)

\bibitem{li2017compactness}
Li, W., Wang, G., Fidon, L., Ourselin, S., Cardoso, M.J., Vercauteren, T.: On
  the compactness, efficiency, and representation of 3{D} convolutional
  networks: brain parcellation as a pretext task. In: IPMI (2017)

\bibitem{pham2017brain}
Pham, C.H., Ducournau, A., Fablet, R., Rousseau, F.: Brain {MRI}
  super-resolution using deep 3{D} convolutional networks. In: ISBI (2017)

\bibitem{pleiss2017memory}
Pleiss, G., Chen, D., Huang, G., Li, T., van~der Maaten, L., Weinberger, K.Q.:
  Memory-efficient implementation of densenets. arXiv:1707.06990  (2017)

\bibitem{plenge2012super}
Plenge, E., Poot, D.H., Bernsen, M., Kotek, G., Houston, G., Wielopolski, P.,
  van~der Weerd, L., Niessen, W.J., Meijering, E.: Super-resolution methods in
  {MRI}: Can they improve the trade-off between resolution, signal-to-noise
  ratio, and acquisition time? Magnetic resonance in medicine  \textbf{68}(6),
  1983--1993 (2012)

\bibitem{sanchez2018brain}
S{\'a}nchez, I., Vilaplana, V.: Brain {MRI} super-resolution using 3{D}
  generative adversarial networks. arXiv preprint arXiv:1812.11440  (2018)

\bibitem{simonyan2014very}
Simonyan, K., Zisserman, A.: Very deep convolutional networks for large-scale
  image recognition. arXiv:1409.1556  (2014)

\bibitem{tustison2010n4itk}
Tustison, N.J., Avants, B.B., Cook, P.A., Zheng, Y., Egan, A., Yushkevich,
  P.A., Gee, J.C.: N4itk: improved n3 bias correction. IEEE transactions on
  medical imaging  \textbf{29}(6), ~1310 (2010)

\bibitem{wang2018recovering}
Wang, X., Yu, K., Dong, C., Change~Loy, C.: Recovering realistic texture in
  image super-resolution by deep spatial feature transform. In: CVPR (2018)

\bibitem{wang2018esrgan}
Wang, X., Yu, K., Wu, S., Gu, J., Liu, Y., Dong, C., Qiao, Y., Loy, C.C.:
  Esrgan: {E}nhanced super-resolution generative adversarial networks. In: ECCV
  (2018)

\bibitem{zhang2018residual}
Zhang, Y., Tian, Y., Kong, Y., Zhong, B., Fu, Y.: Residual dense network for
  image super-resolution. In: CVPR (2018)

\bibitem{zhao2018self}
Zhao, C., Carass, A., Dewey, B.E., Prince, J.L.: Self super-resolution for
  magnetic resonance images using deep networks. In: ISBI (2018)

\end{thebibliography}

\end{document}